\documentclass[aps,prd,11pt,notitlepage,longbibliography,nofootinbib,tightenlines,preprintnumbers]{revtex4-2}
\pdfoutput=1

\makeatletter
\renewcommand{\p@subsection}{}
\renewcommand{\p@subsubsection}{}
\makeatother

\usepackage{amsmath,amssymb,amsfonts}
\usepackage{graphicx}
\usepackage{color}
\usepackage{tikz}
\usepackage{dsfont}
\usepackage[pdftex]{hyperref}
\hypersetup{colorlinks=true, linkcolor=darkred, citecolor=blue, linktoc=page}
\definecolor{darkred}{rgb}{0.8,0.1,0.1}

\definecolor{3dcolor}{rgb}{0.96,0.89,0.76}
\definecolor{4dcolor}{rgb}{0.812,0.851,0.914}

\usepackage{MnSymbol}
\usepackage{physics}
\usepackage{subfigure}

\def\CC{{\mathds{C}}}
\def\RR{{\mathds{R}}}

\newcommand{\mf}{\mathcal{F}}
\newcommand{\trho}{\tilde\rho}
\newcommand{\vrho}{\varrho}

\makeatletter\def\l@subsubsection#1#2{}%
\makeatother

\newcommand{\nocontentsline}[3]{}
\newcommand{\tocless}[2]{\bgroup\let\addcontentsline=\nocontentsline#1{#2}\egroup}

\newcommand{\be}{\begin{equation}}
\newcommand{\ee}{\end{equation}}
\newcommand{\ba}{\begin{array}}
\newcommand{\ea}{\end{array}}
\newcommand{\bea}{\begin{equation}\begin{aligned}}
\newcommand{\eea}{\end{aligned}\end{equation}}

\newcommand{\gym}{g_{\mathrm{YM}}}

\newcommand{\hrho}{\hat{\rho}}

\newcommand{\p}{\partial}

\newcommand{\lam}{\lambda}
\newcommand{\tlam}{\tilde{\lambda}}
\newcommand{\mr}{\mathcal{R}}

\newcommand{\mk}{\mathds{k}}
\newcommand{\exwil}[1]{\ln \left\langle W_{\wedge}(z, #1)\right\rangle}
\newcommand{\deri}[2]{\frac{d #1}{d #2}}
\newcommand{\derit}[2]{\frac{d^2 #1}{d #2^2}}

\begin{document}

\title{Solving $\mathcal N=4$ SYM BCFT matrix models at large $N$}

\author{Dongming He}
\email{dongming.he@vub.be}

\author{Christoph F.~Uhlemann} 
\email{christoph.uhlemann@vub.be}

\affiliation{Theoretische Natuurkunde, Vrije Universiteit Brussel and The International Solvay
	Institutes, Pleinlaan 2, B-1050 Brussels, Belgium}

\begin{abstract}
Many observables in 4d $\mathcal N\,{=}\,4$ SYM with Gaiotto-Witten boundary conditions can be described exactly by matrix models via supersymmetric localization. 
The boundaries typically introduce new degrees of freedom, through a reduction of the gauge symmetry on the boundary or as explicit boundary degrees of freedom, leading to non-trivial matrix models. We derive the saddle points dominating these matrix models at large $N$, expressed in terms of generalized Lambert W-functions.
In string theory the BCFTs are realized by D3-branes ending on D5 and NS5 branes. We independently derive the saddle points from the holographic duals with $\rm AdS_4\times S^2\times S^2\times\Sigma$ geometry and provide precision tests of the dualities.
\end{abstract}

\maketitle
\tableofcontents
\parskip 1mm

\section{Introduction and summary}

Conformal field theories with boundaries (BCFTs) are a rich subject of great interest: realistic physical systems are typically of finite extent, and the presence of boundaries often leads to particularly interesting dynamics. A recent review can be found in \cite{Andrei:2018die}. In string theory BCFTs lead to extended notions of holography \cite{Karch:2000gx,Karch:2000ct,Karch:2022rvr} and have recently found applications in black hole studies. 

In this work we focus on boundaries in 4d $\mathcal N=4$ SYM, the most symmetric and in many ways simplest interacting QFT in 4d. The maximal symmetry that can be preserved in the presence of boundaries comprises 3d Poincar\'e symmetry, enhanced to 3d defect conformal symmetry acting on a half space, and half of the supersymmetries of 4d $\mathcal N=4$ SYM. Such boundary conditions were classified by Gaiotto and Witten \cite{Gaiotto:2008sa,Gaiotto:2008ak}.
The result is a vast space of boundary conditions with intricate S-duality relations between them. They typically introduce new genuinely 3-dimensional degrees of freedom on the boundary, resulting from a reduction of the gauge symmetry on the boundary, couplings to explicit 3d SCFTs, or combinations thereof. The dynamics of these 3d degrees of freedom is crucial for the physics of the entire system.

We study the matrix models describing 4d $\mathcal N=4$ SYM BCFTs after supersymmetric localization.
The boundary degrees of freedom for half-BPS boundaries in 4d $\mathcal N=4$ SYM can be described as 3d SCFTs, denoted $T_\rho^\sigma[SU(N)]$, which arise as strongly-coupled IR limits of 3d quiver gauge theories. The matrix models are constructed by combining matrix models for 4d $\mathcal N=4$ SYM on a hemisphere \cite{Gava:2016oep,KumarGupta:2019nay} with matrix models for 3d $\mathcal N=4$ gauge theories \cite{Kapustin:2009kz,Kapustin:2010xq,Benvenuti:2011ga} using gluing formulas \cite{Dedushenko:2018tgx}. Recent studies include \cite{Wang:2020seq,Komatsu:2020sup,Raamsdonk:2020tin}. Matrix models typically simplify in the large-$N$ or planar limit, where they are often dominated by saddle points, and extensive studies have been performed e.g.\ for standard 4d $\mathcal N=4$ SYM. The presence of boundaries with lower-dimensional degrees of freedom complicates the matrix models, and the natural planar limit for the BCFTs involves quiver theories with a large number of gauge nodes \cite{Coccia:2020wtk,Coccia:2020cku}.
While the saddle points for the 3d theories were found in \cite{Coccia:2020wtk}, following earlier work in \cite{Uhlemann:2019ypp}, the matrix models for the BCFTs are more complicated.

Here we derive the solution for the matrix models for a large class of 4d $\mathcal N=4$ SYM BCFTs in the planar limit, i.e.\ the saddle point eigenvalue distributions dominating the matrix integrals. We provide explicit expressions and show that they solve the saddle point equations.
The solutions involve generalized Lambert W-functions \cite{Corless1996OnTL,2014arXiv1408.3999M} whose inverses are elementary functions.
With the saddle points in hand, insertions into the path integral which do not modify the saddle point at leading order, such as Wilson loops in representations which do not grow too fast and many local operators, can be evaluated immediately, in that sense providing a sort of master field.

Combining non-perturbative field theory methods like supersymmetric localization with string theory and holography often leads to a fruitful interplay. Our results exemplify this.
We derive the saddle points in field theory, with no reference to holography.
But they can also be recovered from string theory, where the BCFTs have holographic duals with $\rm AdS_4\times S^2\times S^2\times \Sigma$ geometry \cite{DHoker:2007zhm,DHoker:2007hhe,Aharony:2011yc,Assel:2011xz}. We extract the saddle points using the results of \cite{Coccia:2021lpp} for Wilson loops, which establish a close connection between the field theory matrix models and certain harmonic functions which define the holographic duals.
This connection is derived and applies to all field theories described by $\rm AdS_4\times S^2\times S^2\times \Sigma$ solutions (for 3d balanced quivers relations to the saddle points of \cite{Coccia:2020wtk} were also explored in \cite{Akhond:2021ffz,Fatemiabhari:2022kpv,Akhond:2022oaf}). 
We make this relation explicit and use it to derive the saddle points for BCFTs from supergravity. 
Combined with the field theory analyses this provides new precision tests of the dualities, which capture e.g.\ the dualities used for black hole studies in \cite{Uhlemann:2021nhu}, and we derive new explicit results for Wilson loops in BCFTs.
The holographic side in turn allows for further generalizations, e.g.\ to ICFTs with fully backreacted D5/NS5 interfaces studied in \cite{Estes:2014hka,Uhlemann:2023oea,Chaney:2024bgx}.

All things combined, the construction covers general 3d $\mathcal N=4$ quiver SCFTs as boundary or interface degrees of freedom, with unbalanced nodes, arbitrary numbers of flavor hypermultiplets attached to an arbitrary number of 3d nodes, and partial Nahm pole boundary conditions. 
For the future it would be interesting to use the saddle points to explore further physical observables, including boundary free energies \cite{Raamsdonk:2020tin}, partition functions e.g.\ on squashed spheres and indices, defect one-point functions \cite{Komatsu:2020sup,Bason:2023bin}, for the full range of theories captured by our analysis.

\subsection{Technical summary}

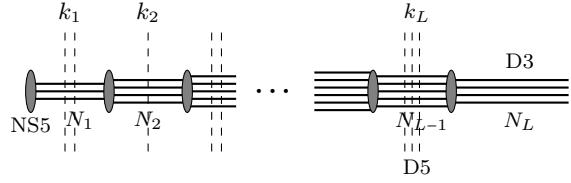
\begin{figure}
	\begin{tikzpicture}[xscale=1.3,yscale=1.0]
		\foreach \i in {-0.1,0,0.1} \draw[thick] (0,\i) -- (0.8,\i);
		\foreach \i in {-0.15,-0.05,0.05,0.15} \draw[thick] (0.8,\i) -- (1.6,\i);
		\foreach \i in {-0.2,-0.1,0,0.1,0.2} \draw[thick] (1.6,\i) -- (2.1,\i);
		
		\node at (2.5,0) {\bf\ldots};	
		
		\foreach \i in {-0.25,-0.15,...,0.26} \draw[thick] (2.9,\i) -- (3.5,\i);
		\foreach \i in {-0.2,-0.1,0,0.1,0.2} \draw[thick] (3.5,\i) -- (4.3,\i);
		
		\foreach \i in {-0.15,-0.05,0.05,0.15} \draw[thick] (4.3,\i) -- (5.5,\i);
		
		\foreach \i in {0,0.8,1.6,3.5,4.3}{ \draw[fill=gray] (\i,0) ellipse (1.5pt and 8pt);}
		
		\foreach \i in {-0.05,0.05} \draw[dashed] (0.4+\i,-0.8) -- +(0,1.6);
		\foreach \i in {0} \draw[dashed] (1.2+\i,-0.8) -- +(0,1.6);
		
		\foreach \i in {-0.075,0,0.075} \draw[dashed] (3.9+\i,-0.8) -- +(0,1.6);
		
		\foreach \i in {-0.05,0.05} \draw[dashed] (1.9+\i,-0.8) -- +(0,1.6);
		%\foreach \i in {-0.05,0.05} \draw (3.1+\i,-0.8) -- +(0,1.6);
		
		\foreach \i in {1,2}{ \node[anchor=south] at ({-0.4+0.8*\i},0.8) {\footnotesize $k_{\i}$};}
		\node[anchor=south] at (3.95,0.8) {\scriptsize $k_{L}$};

		\node at (0.5,-0.38) {\scriptsize $N_1$};
		\node at (1.2,-0.38) {\scriptsize $N_2$};
		\node at (4.0,-0.4) {\scriptsize $N_{L-1}$};   
		\node at (5.0,-0.4) {\scriptsize $N_{L}$};   
		
		\node at (5,0.4) {\scriptsize D3};
		\node at (3.95,-1) {\scriptsize D5};
		\node at (0,-0.45) {\scriptsize NS5};
	\end{tikzpicture}
	\caption{General brane construction, with D3-branes as horizontal lines extending along (0123), NS5 as ellipses along (012456), D5 as vertical dashed lines along (012789). The BCFT emerges in the limit where all NS5-branes become coincident.\label{fig:branes}}
\end{figure}

The class of BCFTs we discuss is engineered in string theory by D3-branes ending on D5 and NS5 branes. In field theory terms they can be described as strongly-coupled IR limit of mixed 3d/4d gauge theories, with 4d $\mathcal N=4$ SYM on a half space coupled to a 3d quiver gauge theory on the boundary.
We use the following notation for the gauge theory,
\begin{align}\label{eq:quiver-gen}
	&U(N_1)-U(N_2)-\ldots - U(N_{L-1})-\widehat{U(N_L)}
	\nonumber\\
	&\hskip 5mm \vert \hskip 15mm \vert \hskip 25mm\vert
	\\
	&\hskip 2mm [k_1] \hskip 9mm [k_2] \hskip 5mm \ldots \hskip 7mm [k_{L-1}]
	\nonumber
\end{align}
The hatted node denotes 4d $\mathcal N=4$ SYM on a half space with gauge group $U(N_L)$, all others are 3d nodes. The dashes denote 3d hypermultiplets in the (anti)fundamental representation of the nodes they connect. $[k_t]$ denotes the flavor symmetry of $k_t$ hypermultiplets in the fundamental representation of a gauge node.
The BCFT emerges as IR limit of this gauge theory and retains the 4d gauge coupling $\gym$ as marginal parameter. 
The brane configuration is illustrated in fig.~\ref{fig:branes}.
When all D5-branes are in the 3d part of the brane construction, as in fig.~\ref{fig:branes}, the entire $SU(N_L)$ part of the 4d $\mathcal N=4$ SYM theory couples to the 3d theory by gauging a global symmetry (with a standard Neumann boundary condition for the $U(1)$ part). 
If some semi-infinite D3-branes terminate on D5-branes, part of the 4d fields satisfies Nahm pole boundary conditions.

Our supergravity derivations capture (partial) Nahm pole boundary conditions. For the field theory discussion we focus on the case where the entire $SU(N_L)$ of 4d $\mathcal N=4$ SYM couples to the 3d SCFT; via S-duality this also gives access to a large class of BCFTs with Nahm pole boundary conditions.
Supersymmetric localization reduces the path integral computing the partition function to a matrix integral with one matrix for each gauge node. This matrix model is
\begin{align}\label{eq:matrix-model-gen}
	\mathcal Z=\,& \frac{1}{N_{1} ! \ldots N_{L} !} \int\left(\prod_{t=1}^{L} \prod_{i=1}^{N_t} d \lambda_{t,i}\right) 
	e^{-\frac{4 \pi^2}{g_{\mathrm{YM}}^2} \sum_{i=1}^{N_L} \lambda_{L, i}^2} \prod_{i<j}^{N_L}\left(\lambda_{L, i}-\lambda_{L, j}\right) 2\sinh\left(\pi\left(\lambda_{L, i}-\lambda_{L, j}\right) \right)
	\nonumber\\
	& \prod_{t=1}^{L-1} \prod_{i<j}^{N_t} 4\sinh^2\left(\pi\left(\lambda_{t,i}-\lambda_{t,j}\right) \right)
	\prod_{t=1}^{L-1} \prod_{i=1}^{N_t} \prod_{j=1}^{N_{t+1}} \frac{1}{2\cosh\left(\pi\left(\lambda_{t,i}-\lambda_{t+1,j}\right)\right)}
	\prod_{t=1}^{L-1} \prod_{i=1}^{N_t} \frac{1}{2^{k_t}\cosh^{k_t}(\pi\lambda_{t,i})}~,
\end{align}
where $\lambda_{t,i}$ are the eigenvalues of the matrix associated with the $t^{\rm th}$ gauge node.
In the integrand the upper line contains the contributions of 4d $\mathcal N=4$ SYM. The lower line contains as first product the contribution from the 3d vector multiplets, then of the bifundamental hypermultiplets connecting adjacent gauge nodes, and lastly from fundamental 3d hypermultiplets.
In the planar limit this matrix integral is dominated by a saddle point, encoded in a family of eigenvalue distributions
\begin{align}\label{eq: continuelimit}
	\tilde\rho_t(\lambda)&=\sum_{i=1}^{N_t} \delta\left(\lambda-\lambda_{t,i}\right)~.
\end{align}
In the planar limit the saddle point eigenvalue distributions can be described by continuous functions. We will show that the saddle points take the form
\begin{align}\label{eq:rho-tilde-cR}
	\tilde\rho_t(\lambda)&=\Im\left[\mathcal R\left(\frac{2\pi\lambda+i\pi t}{L}\right)\right]~,
\end{align}
where $\mathcal R$ is a locally holomorphic function whose form we determine. 
The argument is order one in the planar limit: $t/L$ takes values in the interval $[0,1]$ and the eigenvalues $\lambda$ scale linearly with $L$.\footnote{More precisely, almost all eigenvalues scale linearly with $L$. The largest and smallest eigenvalues grow faster, leading to exponential tails in the eigenvalue distributions which probe brane sources in the holographic duals \cite{Uhlemann:2020bek}.}
In the remainder of this summary we discuss special cases of increasing  generality.

The simplest example is a BCFT where the 3d quiver has all nodes balanced without fundamental hypermultiplets, $k_t=0\,\forall t$. This fixes the quiver to
\begin{align}\label{eq:D3NS5-quiver}
	U(K)-U(2K)-\ldots - U((N_5-1)K)-\widehat{U(N_5K)}
\end{align}
The number of nodes is $L=N_5$, the rank of the 4d $\mathcal N=4$ SYM gauge group $N_L=N_5K$, and the ranks of the 3d nodes are $N_t=t K$. 
This example is engineered by $N_5K$ D3-branes ending on $N_5$ NS5-branes, with $K$ D3-branes ending on each NS5.
For this theory
\begin{align}
	\mathcal R(v)&=\frac{4N_5}{\gym^2} W_k(-e^{v})
	~,
	&
	k&=\frac{\gym^2K}{4N_5}~.
\end{align}
where $W_k$ is a generalized Lambert W-function \cite{Corless1996OnTL,2014arXiv1408.3999M}, defined as solution to the transcendental equation
$z=e^{w}(w+k)/(w-k)\big\vert_{w=W_k(z)}$. It is discussed in app.~\ref{sec:W}, for plots see figs.~\ref{fig:Wk-plot},\ref{fig:Wk-plot-app}.

\begin{figure}
\subfigure[][]{\label{fig:D3NS5-branes}
	\begin{tikzpicture}[xscale=1,yscale=1.2]
		\foreach \i in {-1,0,1} {\draw[thick] (0,{0.05*\i}) -- +(1,0);}
		\foreach \i in {-2.5,-1.5,-0.5,0.5,1.5,2.5} {\draw[thick] (1,{0.05*\i}) -- +(1,0);}
		\foreach \i in {-4,...,4} {\draw[thick] (2,{0.05*\i}) -- +(1,0);}
		\foreach \i in {-5.5,-4.5,...,5.5} {\draw[thick] (3,{0.05*\i}) -- +(1.5,0);}	
				
		\foreach \i in {0,1,2,3}{\draw[fill=gray] (\i,0) ellipse (1.7pt and 8pt);}
		
		\node at (0.5,-0.5) {\scriptsize $K$};
		\node at (1.5,-0.5) {\scriptsize $2K$};
		\node at (2.5,-0.5) {\scriptsize $3K$};
		\node at (3.6,-0.5) {\scriptsize $4K$};
		
		\node at (0,-0.8) { };
	\end{tikzpicture}		
}
\hskip 15mm
\subfigure[][]{\label{fig:D3D5NS5-branes}
		\begin{tikzpicture}[xscale=1,yscale=1.2]
		\foreach \i in {-0.5,0.5} {\draw[thick] (0,{0.07*\i}) -- +(1,0);}
		\foreach \i in {-1.5,...,1.5} {\draw[thick] (1,{0.07*\i}) -- +(1,0);}
		\foreach \i in {-2.5,...,2.5} {\draw[thick] (2,{0.07*\i}) -- +(1,0);}
		\foreach \i in {-2,...,2} {\draw[thick] (3,{0.07*\i}) -- +(1,0);}	
		\foreach \i in {-1.5,...,1.5} {\draw[thick] (4,{0.07*\i}) -- +(1.5,0);}	
		
		\foreach \i in {0,...,4}{\draw[fill=gray] (\i,0) ellipse (1.7pt and 8pt);}
		
		\foreach \i in {-0.15,0,0.15} {\draw[thick,dashed] (2.5+\i,-0.7) -- +(0,1.4);}
				
		\node at (0,-0.8) { };
	\end{tikzpicture}	
}
\caption{Left: Brane construction for D3/NS5, with the quiver in (\ref{eq:D3NS5-quiver}), for $N_5=4$, $K=3$. Right: Brane construction for the quiver in (\ref{eq:D5NS5K-quiver}) with $N_5=5$, $N_{\rm D5}=3$, $R=2$, $S=-2$.\label{fig:branes-2}}
\end{figure}
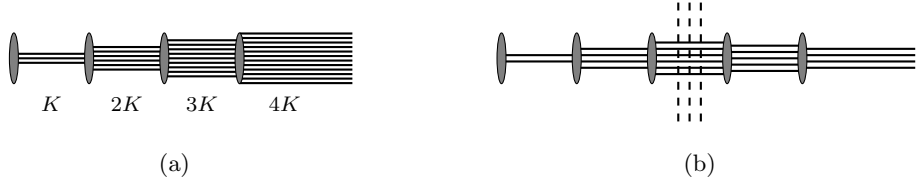

The next example BCFT has a balanced 3d quiver with one group of fundamental flavor multiplets. This fixes the quiver to the following form, with parameters $R,N_5,N_{\rm D5}>0$ and $S<0$,
\begin{align}\label{eq:D5NS5K-quiver}
	U(R)-U(2R)-\ldots &- U(TR) - U(TR-Q)-\ldots - U(N_{\rm D3}+Q) - \widehat{U(N_{\rm D3})}
	\nonumber\\
	&\ \ \ \ \ \ \,\vert\\
	\nonumber & \ \ \ [N_{\rm D5}]	
\end{align}
where $T=N_{5}+S$, $Q=N_{\rm D5}-R$ and $N_{\rm D3}=N_{5}R+N_{\rm D5}S$.
Along the first ellipsis the ranks increase in steps of $R$,
along the second they decrease in steps of $Q$. The total number of gauge nodes is $N_{5}$.
This BCFT is engineered by $N_{\rm D3}$ D3-branes ending on $N_{\rm D5}$ D5-branes and $N_{5}$ NS5-branes in such a way that a net number of $R$ D3-branes end on each NS5 and $S$ D3 on each D5. Both are counted with orientation, and for this quiver $S<0$ (fig.~\ref{fig:D3D5NS5-branes}).
For the saddle point eigenvalue densities it is convenient to express $R$ and $S$ in terms of $K$ and $\delta$ as follows, 
\begin{align}\label{eq:RS-sum}
	R&=K+\frac{2 \cot ^{-1}\!e^{\delta }}{\pi }N_{\rm D5}~, & 
	S&=\frac{\gym^2 K}{4\pi}e^\delta-\frac{2\cot^{-1}\!e^\delta}{\pi}N_{5}~.
\end{align}
Then the saddle point eigenvalue densities are (with $k$ as defined before)
\begin{align}\label{eq:rhot-D3D5NS5-sum}
	\mathcal R(v)&=
	\frac{4N_5}{\gym^2}
	W_k(-e^{v})
	-\frac{i}{\pi} N_{\rm D5}\ln\left(\frac{W_k(-e^{v})-W_k(-e^{i\pi T/N_5})}{W_k(-e^{v})+W_k(- e^{i\pi T/N_5})}\right)
	~.
\end{align}
These two examples illustrate the main features.
For general quivers with 3d fundamental hypermultiplets where all 3d nodes are balanced 
$\mathcal R$ is given by
\begin{align}\label{eq:cR-v-gen-summary}
	\mathcal R(v)&=\frac{4L}{\gym^2}W_k\left(-e^{v}\right)  - \frac{i}{\pi}\sum_{t=1}^{L-1}k_t\ln \left(\frac{W_k\left(-e^{v}\right)-W_k\left(-e^{i\pi z_t}\right)}{W_k\left(-e^{v}\right)-W_k\left(-e^{-i\pi z_t}\right)}\right)~,
\end{align}
with $k$ determined from the normalization of $\tilde\rho$ (eq.~(\ref{eq: normal with flavor})).
The results can be generalized to unbalanced quivers; we sketch the supergravity derivation in sec.~\ref{sec:unbalanced}, the field theory derivations can be generalized following the discussions of unbalanced 3d and 5d quiver SCFTs in \cite{Uhlemann:2019ypp,Coccia:2020wtk}.

\medskip
\textbf{Outline:}
In sec.~\ref{sec:matrix} we discuss the BCFT matrix models in the continuum quiver formulation of \cite{Uhlemann:2019ypp} and solve the saddle point equations. 
In sec.~\ref{sec:sugra} we give the general prescription for deriving the saddle points from the supergravity duals, which extends e.g.\ to unbalanced quivers and partial Nahm pole boundary conditions, and apply it to the aforementioned examples.
In sec.~\ref{sec:prec-text} we spell out Wilson loop expectation values and how they provide precision tests of the dualities.

\section{Matrix models in continuum formulation}\label{sec:matrix}

We rewrite the matrix model (\ref{eq:matrix-model-gen}) in the continuum formulation for long quivers, following  \cite{Uhlemann:2019ypp,Coccia:2020wtk}. To this end we denote the integrand in (\ref{eq:matrix-model-gen}) by $\mf$, defined as follows,
\begin{align}
\mathcal Z&= \frac{1}{N_{1} ! \ldots N_{L} !} \int\left(\prod_{t=1}^{L} \prod_{i=1}^{N_t} d \lambda_{t,i}\right)\, e^{-\mf}~.
\end{align}
To transition to the continuum formulation, we use the eigenvalue densities in \eqref{eq: continuelimit} and in addition introduce
\begin{align}
z&=\frac{t}{L}~, 
&
\tilde \rho_t(\lambda)&\rightarrow \tilde\rho(z,\lambda)~,
&
\lbrace N_t\rbrace &\rightarrow N(z)~,
&
k(z)=\sum_{t=1}^L \frac{k_t}{L} \delta\left(z-z_t\right)~,
\end{align}
such that $z$ is an effectively continuous coordinate along the quiver, the ranks are encoded in an effectively continuous function $N(z)$ and the eigenvalue densities in a function of two real variables. This further leads to the replacement of sums by integrals, $\sum_{t=1}^L f_t \rightarrow L \int_0^1 d z f(z L)$.
The rewriting of the matrix models follows the steps in \cite{Uhlemann:2019ypp,Coccia:2020wtk}.
We split the result as
\begin{align}
\mf&=\mf_i+\mf_\partial~,
\end{align}
where $\mf_i$ contains integrals over $z$ while $\mf_\partial$ encodes boundary terms.
They are given by 
\begin{align}
	\mf_i=\,&\int\! dz\! \int d \tlam d \lam\Big[-L \trho(z, \tlam) \trho(z, \lam)\, \ln\tanh\left(\pi|\tlam-\lam| \right)
	- \frac{\partial_z\trho(z, \tlam) \partial_z \trho(z, \lam)}{2L}\ln \left[2\cosh\left (\pi(\tlam-\lam)\right) \right]\Big]
	\nonumber\\
	&
	+L\! \int \! dz\,k(z) \!\int\! d\lam\, \trho(z,\lam)\ln \left[2\cosh(\pi \lam)\right]\,,
	\nonumber\\
	\mf_\partial=\,&\frac{4 \pi^2}{g_{\mathrm{YM}}^2} \int d \lam\, \lam^2 \trho(1,\lam)-\frac{1}{2} \int d \tlam d \lam\,\trho(1,\tlam) \trho(1,\lam)\, \ln |\tlam-\lam| 
	\nonumber\\
	&-\frac{1}{2} \int \!\!d \tlam d \lam\left[ \trho(1,\tlam) \trho(1,\lam)\, \ln \tanh\left(\pi|\tlam-\lam| \right)
	+\trho(0,\tlam) \trho(0,\lam)\, \ln \left[2\cosh\left (\pi(\tlam-\lam)\right)\right]\right]\,,
\end{align}
where integrals with divergent integrands, such as $\ln \abs{\lam-\tlam}$, are understood as principle value integrals.
The scaling analysis for the eigenvalues based on the bulk part of $\mf$ is unchanged compared to \cite{Uhlemann:2019ypp,Coccia:2020wtk}, leading to linear scaling with $L$. We thus set
\begin{align}
\lambda&=Lx~, & 
\vrho(z,x)&=L\trho(z,\lambda)~.
\end{align}
With the following approximations for large arguments,
\begin{align}
	\ln\tanh|x| &\approx -\frac{1}{4} \delta(x)~, 
	&
	\ln \left[2\cosh(x)\right] &\approx |x|~,
\end{align}
we then arrive at
\begin{align}
	\mf_i=\,&\frac{\pi}{4} \int dz \left[\int d x \vrho^2(z, x)-2\int dx dy|x-y| \partial_z \vrho(z, x) \partial_z \vrho(z, y)+4L^2\int dx \vrho(z,x) k(z) |x|\right]~,
 \nonumber\\
 \mf_{\p}= \,& \frac{4 \pi^2 L^2}{\gym^2} \int d y\, y^2 \vrho(1, y)-\frac{1}{2} \int d x d y \ln \abs{x-y} \vrho(1, x) \vrho(1, y)+\frac{\pi}{8L} \int d x \vrho(1, x)^2 
 \nonumber\\
& -\frac{\pi L}{2}\int d x d y \abs{x-y} \vrho(0, x) \vrho(0, y)-\frac{N(1)^2\ln L }{2}~.
\end{align}
This is the starting point for the discussion of saddle points and the connection to supergravity.

\subsection{Saddle point equations and solutions}

The saddle point conditions are obtained by demanding variations of $\mf$ to vanish. 
The bulk part is unmodified compared to the 3d analysis in \cite{Coccia:2020wtk}, leading for $0<z<1$ to the condition
\begin{equation}\label{eq:saddle-bulk}
	\partial^2_ x \vrho(z, x)+4 \partial ^2_z \vrho(z, x)+4L^2 \delta(x)k(z)=0~.
\end{equation}
The boundary conditions arise from a combination of the variations of $\mf_{\p}$ and boundary terms from integration by parts when varying $\mf_i$. They are given by
\begin{align}
z&=0: & \frac{1}{L} \int d y  \abs{x-y} \partial_z \vrho(0, y)- \int d y \abs{x-y} \vrho(0, y)&=0~,
\label{eq: z=0 initial}
\\
z&=1: & -\pi\int d y\, \abs{x-y}\, \partial_z \vrho(z, y)\big|_{z=1}+\frac{4 \pi^2 L^2}{\gym^2 } x^2-\int d y\,  \vrho(1, y)\ln \abs{x-y}+\frac{\pi}{4L} \vrho(1, x)&=0~.
\label{eq: z=1 initial}
\end{align}
We note that $\varrho$ is order $N(z)$.
Both conditions simplify in the planar limit $L\rightarrow\infty$,
which leads to a Dirichlet boundary condition at $z=0$
\begin{align}
\vrho(0,x)=0~.
\end{align}
This is the condition associated with quiver tails of decreasing-rank gauge nodes derived in \cite{Uhlemann:2019ypp,Coccia:2020wtk}.
At $z=1$, the last term in \eqref{eq: z=1 initial} is subleading. Taking a derivative with respect to $x$ leads to
\be\label{eq: z=1}
-\pi\int_{-x}^x  dy~ \p_z\vrho(z,y)\big\vert_{z=1}+\frac{8\pi^2L^2}{\gym^2}x-\int dy\frac{\vrho(1,y)}{x-y}=0~.
\ee
These are the complete saddle point equations.
The difference to 3d SCFTs lies in the boundary condition at $z=1$.
Without the first term the latter would be the saddle point equation for 4d $\mathcal N=4$ SYM. 
The derivative term couples the eigenvalues of the 4d node to those of the 3d quiver. 

The bulk saddle point equation (\ref{eq:saddle-bulk}) implies that $\vrho$ is locally a harmonic function of the complex combination $v=2\pi x+i\pi z$, with appropriate singularities at the locations $v=i\pi z_t$ of flavors. It can thus be written as imaginary part of a locally holomorphic function as in (\ref{eq:rho-tilde-cR}),
\begin{align}\label{eq:varrho-cR}
\vrho&=L\Im(\mathcal R(v))~, & v&=2\pi x+i\pi z~.
\end{align}
The equations are invariant under $x\rightarrow -x$ and the saddle points will satisfy $\vrho(z,-x)=\vrho(z,x)$. This translates to
\begin{align}\label{eq:calR-conj}
\overline{\mathcal R(-\bar v)}&=-\mathcal R(v)~.
\end{align}
The boundary conditions are $\vrho\vert_{\Im(v)=0}=0$ and (\ref{eq: z=1}), which we may write using the Cauchy-Riemann equations and the symmetry of $\mathcal R$ as
\begin{align} \label{eq: saddle in R}
-\pi\Re\mathcal R(2\pi x+i\pi)
+\frac{8\pi^2 L}{\gym^2} x - \int dy \frac{\Im\mathcal R(2\pi y+i\pi)}{x-y}&=0~.
\end{align}
We will show that the solution is
\begin{align}\label{eq: saddle field theory}
\mathcal R(v)&=\frac{4L}{\gym^2}W_k\left(-e^{v}\right)  - \frac{i}{\pi}\sum_{t=1}^{L-1}k_t\ln \left(\frac{W_k\left(-e^{v}\right)-W_k\left(-e^{i\pi z_t}\right)}{W_k\left(-e^{v}\right)-W_k\left(-e^{-i\pi z_t}\right)}\right)~,
\end{align}
where $W_k$ is the generalized Lambert W-function discussed in app.~\ref{sec:W}, with $k$ to be determined from normalization. This satisfies the bulk saddle point equation (\ref{eq:saddle-bulk}) with the correct sources  and with mirror charges to implement the boundary condition at $z=0$, leaving only (\ref{eq: saddle in R}) to be verified.

\subsection{Derivation of the solutions}

Now we derive this solution and demonstrate its validity. First we consider the case without flavors, with the quiver diagram (\ref{eq:D3NS5-quiver}). For this theory $L=N_5$ and the rank function is
\begin{align}
N(z)=K N_5 z~.
\end{align}
As discussed in (\ref{eq:varrho-cR}), $\vrho$ is locally the imaginary part of a holomorphic function $\mathcal R$ on the strip parametrized by the complex coordinate $v$. This means we can use $\mathcal R$ as complex coordinate in the interior of the strip. We construct $\mathcal R$ through the mapping back to the $v$ coordinate,  $v(\mathcal R)$.

We first note that $\varrho(z,x)$ should have compact support at $z=1$: When $x$ is large in  (\ref{eq: z=1}), the second term is unbounded while the other two are bounded. The equation has to be imposed on the support of $\varrho(1,x)$, so making the support finite avoids a contradiction.
We take
\begin{align}\label{eq:rho-supp}
	&v=2\pi x+i\pi\,, \ |x|>x_0 \qquad \Rightarrow \qquad \Im\left(\mathcal R(v)\right)=0~,
\end{align}
with $x_0$ to be determined. The region outside the support at $z=1$ is mapped to the real line in the $\mathcal R$ coordinate, while the support is mapped away from the real axis.
The region $z=0$ also maps to the real line in the $\mathcal R$ coordinate.
This leads to the ansatz
\begin{align}\label{eq:v-ansatz}
	v(\mathcal R)&=v_0(\mathcal R)- \ln\left(\frac{r_1-\mathcal R}{\mathcal R-r_0}\right)~,
\end{align}
with real $r_1>r_0$ and $v_0(\mathcal R)$ to be determined. If $v_0(\overline{\mathcal R})=\overline{v_0(\mathcal R)}$, so that $v_0(\mathcal R)$ is real for real $\mathcal R$, 
this maps $\mathcal R>r_1$ and $\mathcal R<r_0$ to $v=i\pi + x$ with real $x$, while $r_0<\mathcal R<r_1$ maps to real $v$. In view of (\ref{eq:calR-conj}) we have $-r_0=r_1\equiv r$.
We note that $v$ with large positive $x$ maps to $\mathcal R\sim r$, while $v$ with large negative $x$ maps to $\mathcal R\sim -r$.
It remains to determine $r$ and $v_0$.

We now turn to (\ref{eq: saddle in R}) itself. We set $v=2\pi x+i\pi$ and $\hat v=2\pi y+i\pi$,
and denote by $\mathcal C$ the contour comprising the line $\Im(\hat v)=\pi$. 
This leads to
\begin{align}\label{eq:saddle-tmp2}
	-\pi\Re\mathcal R(v)
	+\frac{4\pi L}{\gym^2} \Re(v) - \int_{\mathcal C} d\hat v \frac{\Im\mathcal R(\hat v)}{v-\hat v}&=0~.
\end{align}
As discussed above, $\mathcal R(v)$ is real when $z=1$ with $|x|>x_0$, so the integrand has compact support.
We assume that $\mathcal R$ has at most a simple pole at $v=\pm\infty+i\pi$, i.e.\ up to regular terms
\begin{align}
	\mathcal R(v)&\sim\beta v~.
\end{align}
We can then take the imaginary part out of the integral in (\ref{eq:saddle-tmp2}), producing a principal value integral along a closed contour on the Riemann sphere with poles at $\hat v_1=v$ and $\hat v_2=\infty+i\pi$ on the contour,
\begin{align}\label{eq: p.v.contour}
	\int_{\mathcal C} d\hat v \frac{\Im\mathcal R(\hat v)}{v-\hat v}&=
	\frac{1}{2i}\int_{\mathcal C} d\hat v \frac{\mathcal R(\hat v+ i\epsilon)}{v-\hat v}+{\rm c.c.}
	\nonumber\\
	&=
	\frac{1}{2i}\left[\int_{\mathcal C_0} d\hat v \frac{i\mathcal R(\hat v+ i\epsilon)}{v-\hat v}
	-i\pi \mathcal R(\hat v_1+i\epsilon)
	-i\pi \mathcal R(\hat v_2+i\epsilon)\right]+{\rm c.c.}
\end{align}
In the first line part of the extended integral traces the branch cuts of $\mathcal R(v)$, and the $\epsilon$-prescription selects the branch so that $\mathcal R$ is holomorphic inside the contour ($\Im(\hat v)>\pi$).
In the second line the Sokhotski–Plemelj formula is applied: $\mathcal C_0$ is a modified contour with the poles outside and the residues compensate for the change.
The integral along $\mathcal C_0$ vanishes.
The residue at $\hat v_1$ cancels the first term in (\ref{eq:saddle-tmp2}). The residue at $\hat v_2$ cancels the second term in (\ref{eq:saddle-tmp2}) if $\beta=4L/\gym^2$.
The last remaining equation is then satisfied and we arrive at
\begin{align}\label{eq:cR-no-flavor}
	v(\mathcal R)&=\frac{\mathcal R}{\beta}- \ln\left(\frac{r-\mathcal R}{\mathcal R+r}\right)
	\qquad
	\longleftrightarrow
	\qquad
	\mathcal R(v)=\beta W_{r/\beta}\left(-e^v\right)~,
	&
	\beta&=\frac{4L}{\gym^2}~.
\end{align}
This has the desired features (see fig.~\ref{fig:Wk-plot}) and the W-function is discussed in appendix \ref{sec:W}.

The last step is to determine $r$ from from the requirement that $\varrho$ should be normalized to $N(z)$.
It is enough to impose this at $z=1$, as the remaining eigenvalue distributions are fixed once this condition is set.
This integration can be carried out following the strategy used in \eqref{eq: p.v.contour}, with the result determined by the residue at $\hat v=\infty+i\pi+i\epsilon$,\footnote{To evaluate the residue at $v=\infty+i(\pi+\epsilon)$, corresponding to $\mr=\infty$, one can use that $v(\mr)$ is simpler than $\mr(v)$ and change the parametrization of the integrand from $v$ to $\mr$ via $ \mr(v)dv=\mr\frac{dv}{d\mr}d\mr=\mr(\beta^{-1}+2r/(r^2+\mr^2))d\mr$.}
\begin{align}\label{eq:no-flavor-rho-norm}
	N(1)=LK&=\int dy~\vrho(1,y)=-\frac{iL}{2}\int\frac{dv}{2\pi}\mr(v)+{\rm c.c}=L r~.
\end{align}
We conclude that $r=K$ and
\begin{align}
\frac{r}{\beta}&=\frac{\gym^2K}{4L} \equiv k~,
&
\mathcal R(v)&=\frac{4L}{\gym^2} W_k(-e^v)~.
\end{align}
This means
$\vrho(1,x)$ has compact support at $|x|<x_0$ with $2\pi x_0=\sqrt{k(k+2)}+\ln(k+1+\sqrt{k(k+2)})$. 
We validated this solution by numerically verifying that it solves the saddle point equation (\ref{eq: z=1}). 
We also compared to a numerical study of the matrix model (\ref{eq:matrix-model-gen}) in which we solved the saddle point equations for finite $L$ and $K$ using the method of \cite{Fluder:2018chf} and found qualitative agreement.\footnote{The match to string theory calculations in sections \ref{sec:sugra} and \ref{sec:prec-text} provides an independent and analytic validation.}

The derivation can be extended to general balanced quivers with flavors.
Implementing compact support at $z=1$ again leads to the ansatz (\ref{eq:v-ansatz}).
However, the bulk saddle point equation (\ref{eq:saddle-bulk}) now has sources.
This leads to
\begin{align}
	\mathcal R(v)&=\mathcal R_0(v) + \mathcal R_f(v)~,
\end{align}
where $\mathcal R_0(v)$ is given by (\ref{eq:cR-no-flavor}) with $\beta$ unchanged and $r$ as free parameter.
With $r/\beta\equiv k$ we have
\begin{align}\label{eq: green source}
\mathcal R_0(v)&=\frac{4L}{\gym^2} W_k(-e^v)~,
&
\mathcal R_{f}(v)&=-\frac{i}{\pi}\sum_{t=1}^{L-1}k_t\ln \left(\frac{W_k\left(-e^{v}\right)-W_k\left(-e^{i\pi z_t}\right)}{W_k\left(-e^{v}\right)-W_k\left(-e^{-i\pi z_t}\right)}\right)~,
\end{align}
where $\mr_f$ follows from the Green's function on the half plane with Dirichlet boundary condition.
Imposing the correct normalization again determines $r$. In analogy to \eqref{eq:no-flavor-rho-norm},
\begin{align}\label{eq: normal with flavor}
	N(1)&=-\frac{iL}{2}\int\frac{dv}{2\pi}\mr(v)+{\rm c.c}
	=Lr
	-\left[\frac{iL}{2}\int\frac{dv}{2\pi}\mr_f(v)+{\rm c.c}\right]
	=Lr-\frac{i L}{\pi}\sum_{t=1}^{L-1}k_t W_k(-e^{i\pi z_t})~.
\end{align}
This determines $r$ (and $k$) in terms of the parameters of the field theory under consideration.

As an example we discuss the theory (\ref{eq:D5NS5K-quiver}), where $N(1)=N_{\rm D3}=N_5R+N_{\rm D5}S$. The location of the only flavor is $z_t=T/L=1+S/L$, and with the expressions in (\ref{eq:RS-sum}) we have $W_k(-e^{i\pi z_t})=i e^{\delta}k$, with $k$ defined as $k=\gym^2 K/(4N_5)$. With the expressions for $R$, $S$ in (\ref{eq:RS-sum}), (\ref{eq: normal with flavor}) is satisfied with $r=K$.

In closing we point out that although the eigenvalue distributions are given in terms of a special function, they can be expressed in terms of elementary functions in parametrized form with complex coordinate $\mr_0$, by setting $2\pi x+i\pi z=\mr_0/\beta-\ln\left[(r-\mr_0)/(\mr_0+r)\right]$ and $\vrho(\mr_0)=\mr_0+\mr_f(\mr_0)$.

\section{Saddle points from supergravity}\label{sec:sugra}

The general supergravity solutions describing the near-horizon limit of D3, D5 and NS5 branes intersecting at a 3d locus were constructed in \cite{DHoker:2007zhm,DHoker:2007hhe}. The geometry, $\rm AdS_4\times S^2\times S^2\times\Sigma$, is a warped product of $\rm AdS_4$ and two spheres over a Riemann surface $\Sigma$.
Concrete solutions are specified by a pair of locally harmonic functions $h_{1/2}$ on $\Sigma$. Any choice of $(\Sigma,h_{1/2})$ gives a supersymmetric solution to the equations of motion. The explicit expressions for metric, dilaton and remaining Type IIB fields will not be needed here, they are in \cite{DHoker:2007zhm,DHoker:2007hhe} (a concise summary is in \cite[sec.~4.1]{Coccia:2021lpp}).

Physically sensible solutions are obtained by imposing regularity conditions on the 10d solutions. This in general constrains the admissible choices for $h_{1/2}$ up to a number of parameters which match the parameters in the brane construction. This was worked out for the BCFTs of interest here in \cite{Aharony:2011yc} and for 3d linear and circular quivers in \cite{Assel:2011xz,Assel:2012cj}.

\begin{figure}
	\subfigure[][]{\label{fig:sugra-gen}
	\begin{tikzpicture}[xscale=0.9,yscale=1.1]
		\shade [right color=3dcolor!100,left color=3dcolor!100] (-0.3,0)  rectangle (0.3,-2);
		
		\shade [ left color=3dcolor! 100, right color=4dcolor! 100] (0.3-0.01,0)  rectangle (2,-2);
		\shade [ right color=3dcolor! 100, left color=3dcolor! 100] (-0.3+0.01,0)  rectangle (-2,-2);
		
		\draw[thick] (-2,0) -- (2,0);
		\draw[thick] (-2,-2) -- (2,-2);
		\draw[dashed] (2,-2) -- +(0,2);
		\draw[thick] (-2,-2) -- +(0,2);
		
		\node at (-1,-0.5) {$\Sigma$};
		\node at (2.6,-0.65) {\footnotesize $\rm AdS_5$};
		\node at (2.6,-1) {\footnotesize $\times$};
		\node at (2.6,-1.35) {\footnotesize $\rm S^5$};

		\foreach \i in {-1.1,-0.5,1}{
			\draw[very thick] (1.2*\i+0.1,-0.08) -- (1.2*\i+0.1,0.08) node [anchor=south] {\footnotesize D5};}
		\node at (0.3,0.3) {\footnotesize $\cdots$};
		\foreach \i in {-1.2,-0.3,1.1}{
			\draw[thick] (0.9*\i-0.1,-1.92) -- (0.9*\i-0.1,-2.08) node [anchor=north] {\footnotesize NS5};
		}			
        \node at (0.3,-2.3) {\footnotesize $\cdots$};
	\end{tikzpicture}
	}
	\hskip 10mm
	\subfigure[][]{\label{fig:Sigma-h2D-contours}
		\begin{tikzpicture}
			\node at (0,0) {\includegraphics[width=0.25\linewidth]{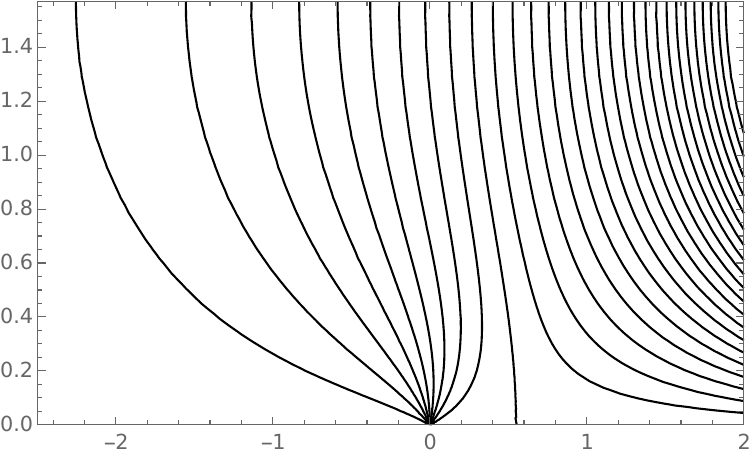}};
			\draw[thick] (0.3,-1.08-0.08) node [anchor=north,yshift=0.5mm] {\scriptsize NS5} -- +(0,0.16);
			\node at (-1,0.5) {$\Sigma$};
		\end{tikzpicture}
	}
	\caption{Left: Schematic form of $\rm AdS_4\times S^2\times S^2\times\Sigma$ solutions dual to BCFTs. For balanced 3d quivers there is only one NS5 source on the lower boundary. Right: Curves of constant $h_2^D$ for $h_2$ in (\ref{eq:h1h2-D3NS5-BCFT-p}). The curves starting on the lower boundary describe D5-branes which hololographically realize Wilson loops.}
\end{figure}
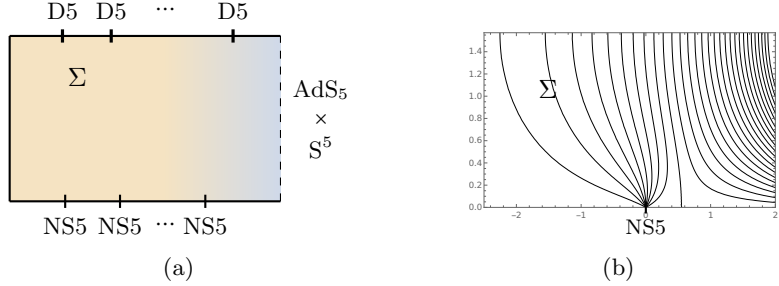

We focus on BCFTs arising from gauge theories of the form (\ref{eq:quiver-gen}), as engineered by the brane setups in figs.~\ref{fig:branes}, \ref{fig:branes-2}. 
A schematic illustration of the supergravity solutions is in fig.~\ref{fig:sugra-gen}. Each D5 source represents a group of flavor hypermultiplets in the quiver (\ref{eq:quiver-gen}), and each NS5 source represents a string of balanced gauge nodes. For these solutions $\Sigma$ can be taken as a strip
\begin{align}\label{eq:Sigma-z}
	\Sigma=\left\lbrace z\in\CC \,\big\vert\, 0\leq\Im(z)\leq \frac{\pi}{2}\right\rbrace~.
\end{align}
We start with BCFTs of the form (\ref{eq:quiver-gen}) with all 3d nodes balanced. The generalization to unbalanced quivers will be discussed in sec.~\ref{sec:unbalanced}.
For balanced BCFTs,
\begin{align}\label{eq:h1h2-D3NS5-BCFT-p}
	h_1&=-\frac{i\pi\alpha^\prime}{4} K e^z-\frac{\alpha^\prime}{4}\sum_{p=1}^P  N_{\rm D5}^{(p)}\ln\tanh\left(\frac{i\pi}{4}-\frac{z-\delta_p}{2}\right) +{\rm c.c.}
	\nonumber\\
	h_2&=\frac{\pi\alpha^\prime}{4}K e^{z+2\phi_0}-\frac{\alpha^\prime}{4}N_5\ln\tanh\left(\frac{z}{2}\right) +{\rm c.c.}
\end{align}
The exponential terms give rise to a locally $\rm AdS_5\times S^5$ region at $\Re(z)\rightarrow\infty$ with dilaton set by $\phi_0$. This produces an $\rm AdS_4$, which is conformally equivalent to a half space, as conformal boundary.
The pole in $\partial h_2$ at $z=0$ represents a single group of $N_5$ NS5-branes, while the poles in $\partial h_1$ at $z=\delta_p+\frac{i\pi}{2}$ represent D5-brane groups. The residues determine the 5-brane charge for each group, the locations determine how many D3-branes end on the group of 5-branes (see \cite[sec.~4]{Assel:2011xz}).

In app.~\ref{app:sugra-deriv} we derive an explicit relation between the harmonic functions $h_{1/2}$ and the saddle point eigenvalue densities $\tilde \rho_t$ for the matrix model associated with the dual field theory. The relation follows from the expression for antisymmetric Wilson loop expectation values, represented by additional D5-branes in the brane configuration, obtained in \cite{Coccia:2021lpp}. 
It yields a parametrized form
\begin{align}\label{eq:b-rho-h12}
\lambda&=\frac{h_2}{\pi \alpha'}~,
&
t&=\frac{2 |h_2^D|}{\pi\alpha'}~,
&
\tilde\rho_t(\lambda) &=\frac{2h_1}{\pi \alpha'}~,
\end{align}
where $h_2^D$ is the dual harmonic function, defined up to a shift by $h_2^D=i(A_2-\overline{\mathcal A}_2)$ after splitting $h_2$ into holomorphic and anti-holomorphic parts as $h_2=\mathcal A_2+\overline{\mathcal A}_2$.
The field theory quiver coordinate is given by $t$ up to a possible shift and reflection to match conventions.
Upon tracing through a part of $\Sigma$ which hosts Wilson loop D5-branes (fig.~\ref{fig:Sigma-h2D-contours}), this reconstructs the full eigenvalue densities.\footnote{We note that $h_2$ is non-negative on $\Sigma$, which may appear to only cover $\lambda\geq 0$. As noted in app.~\ref{app:sugra-deriv}, however, the Wilson loop D5-branes explore a double cover of $\Sigma$ and reconstruct the full densities. This can be implemented by extending $h_2$ to $0<\Im(z)<\pi$. Alternatively, the densities for $\lambda < 0$ follow from symmetry under $\lambda\rightarrow -\lambda$.\label{foot:ext}}
We note that these relations are reminiscent of relations observed in a different context in \cite{Okuda:2008px}.

\subsection{Solving for \texorpdfstring{$\rho$}{rho}}
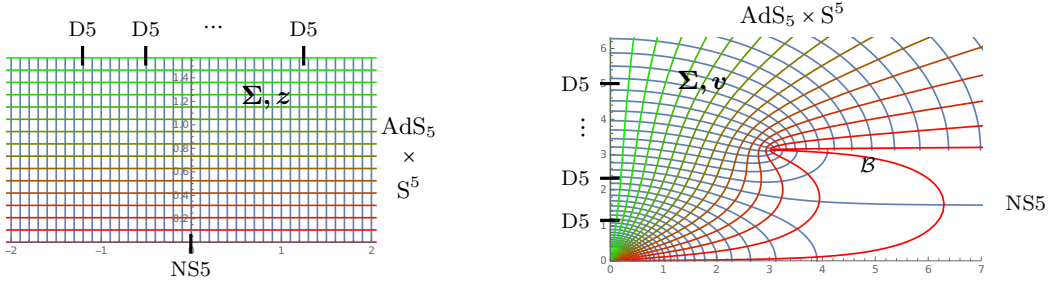
\begin{figure}
	\begin{tikzpicture}
		\node at (-8,0) {\includegraphics[width=0.3\linewidth]{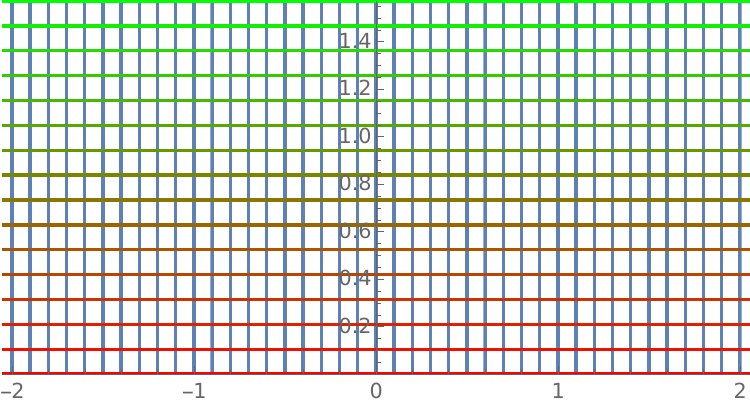}};
		\node at (-7,0.8) {{\boldmath{$\Sigma,z$}}};
		
		\draw[very thick] (-8,-1.15+0.13) -- + (0,-0.26) node [yshift=-2mm] {\scriptsize NS5};
		\foreach \i in {-1.1,-0.5,1}{
			\draw[very thick] (-8+1.4*\i+0.1,1.35-0.13) -- (-8+1.4*\i+0.1,1.35+0.13) node [anchor=south] {\scriptsize D5};}
		\node at (-8+0.3,1.7) {\footnotesize $\cdots$};
		
		\node at (-8+2.9,+0.4) {\footnotesize $\rm AdS_5$};
		\node at (-8+2.9,0) {\footnotesize $\times$};
		\node at (-8+2.9,-0.4) {\footnotesize $\rm S^5$};

		\node at (0,0) {\includegraphics[width=0.31\linewidth]{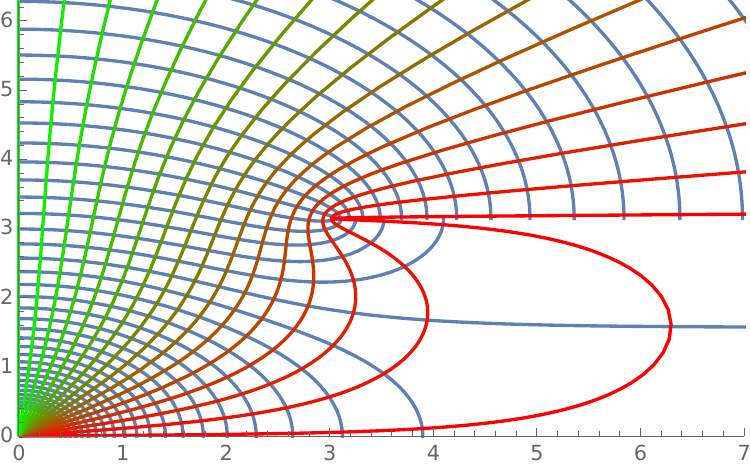}};
		\node at (-1.2,1) {{\boldmath{$\Sigma,v$}}};
		\node at (0,1.9) {\footnotesize $\rm AdS_5\times S^5$};
		\foreach \i in {-0.6,-0.2,0.7}{
			\draw[very thick] (-2.3,1.4*\i) -- +(-0.26,0) node [anchor=east] {\scriptsize D5};}
		\node [rotate=90] at (-2.75,0.4) {$\cdots$};
		\node at (1,-0.1) {\scriptsize $\mathcal B$};
		\node at (3.1,-0.6) {\scriptsize NS5};
	\end{tikzpicture}
	\caption{$\Sigma$ as strip and as upper right quadrant in $v$ coordinate (\ref{eq:v-coord}). 
	The boundary $\Im(z)=\frac{\pi}{2}$ maps to the positive imaginary axis in the $v$ coordinate. 
	The boundary segment $z<0$ of the real line maps to the positive real axis $v\in\RR_+$. The boundary segment $z>0$ maps to a double cover of $\mathcal B$ in (\ref{eq:cB}) in the $v$ coordinate; the lines $\Im(z)=\epsilon$ map one-to-one to curves which `hug' $\mathcal B$.
	The NS5 source at $z=0$ maps to $\Re(v)=+\infty$ with $0<\Im(v)<\pi$. The $\rm AdS_5\times S^5$ region at $\Re(z)=\infty$ maps to $|v|=\infty$ with $\Im(v)>\pi$.\label{fig:BCFT-coord-transf}}
\end{figure}

To obtain a more explicit form we can choose $h_2+h_2^D$ as complex coordinate on $\Sigma$. We now make this explicit and discuss the global structure. We introduce a new coordinate $v$ defined by
\begin{align}\label{eq:v-coord}
	ke^z-\ln\tanh\frac{z}{2}+i\pi&=v
	&&\leftrightarrow
	&
	z&=\ln\left(\frac{W_k(-e^{v})}{k}\right)~,
	&
	k\equiv \frac{\pi Ke^{2\phi_0}}{N_5}~,
\end{align}
where $W_k$ is the generalized Lambert W-function discussed in app.~\ref{sec:W}. The coordinate transformation specifies the relevant branch. In this $v$ coordinate
\begin{align}
	h_1&=-\frac{i\alpha'}{4}N_5e^{-2\phi_0} W_k(e^{-v})
	-\frac{\alpha^\prime}{4}\sum_{p=1}^P N_{\rm D5}^{(p)}\ln\left(\frac{W_k(-e^{v})-ik e^{\delta_p}}{W_k(-e^{v})+ike^{\delta_p}}\right)+{\rm c.c.}
	~,
	\nonumber\\
	h_2&=\frac{\alpha'}{4}N_5(v+\bar v)~, \qquad h_2^D=\frac{i\alpha'}{4}N_5(v-\bar v)~,
\end{align}
so that $v$ directly relates to a complex combination of $\lambda$ and $t$ via (\ref{eq:b-rho-h12}).
The coordinate transformation from $z$ to $v$ maps the strip in (\ref{eq:Sigma-z}) to the upper right quadrant with a branch cut along
\begin{align}\label{eq:cB}
	\mathcal B&=\left\lbrace v=x+i\pi\,\big\vert\, x>\sqrt{k(k+2)}+\ln(k+1+\sqrt{k(k+2)})\right\rbrace~.
\end{align}
The boundary segment $\Im(z)=0$ maps to this branch cut; in the interior of $\Sigma$ the map is one-to-one.
For more details see fig.~\ref{fig:BCFT-coord-transf}.
Extending $h_2$ to a double cover of $\Sigma$ (see footnote \ref{foot:ext}) to cover negative $\lambda$ extends the right plot in fig.~\ref{fig:BCFT-coord-transf} to the full upper half plane by reflection.

Using (\ref{eq:b-rho-h12}) we can now translate the harmonic functions specifying the holographic dual to an explicit expression for the saddle point eigenvalue distributions
\begin{align}\label{eq:rho-t-sugra-balanced}
	\tilde\rho_t(\lambda)&=
	\frac{1}{\pi}\Im\left[
	N_5e^{-2\phi_0} W_k(-e^{v})
	-i\sum_{p=1}^P N_{\rm D5}^{(p)}\ln\left(\frac{W_k(-e^{v})-ik e^{\delta_p}}{W_k(-e^{v})+ike^{\delta_p}}\right)
	\right]
	~,
	&
	v&=\frac{2\pi|\lambda|+i\pi t}{N_5}~.
\end{align}
For the translation to field theory quantities we take into account the dilaton convention of \cite{DHoker:2007zhm,DHoker:2007hhe}, in which $\tau=\chi+ie^{-2\phi}$, for extracting the 4d gauge coupling and arrive at
\begin{align}
\gym^2&=4\pi e^{2\phi_0}~, &
k_t&=N_{\rm D5}^{(t)}~, &
L&=N_5~.
\end{align}
It remains to express $\delta_p$ in terms of field theory parameters. To this end, we note that the D3-brane charge at a D5-brane source representing flavors is given by \cite{Assel:2011xz}
\begin{align}
	N_{\rm D3}^{(p)}&=-\frac{2 h_2^D(z=\delta_p+\frac{i\pi}{2})}{\pi \alpha'} N_{\rm D5}^{(p)}
	=-\frac{iL}{\pi}\left[i e^{\delta_p}k+\ln \frac{i e^{\delta_p}+1}{i e^{\delta_p}-1}\right]N_{\rm D5}^{(p)}~,
\end{align}
where we used $h_2^D$ resulting from (\ref{eq:h1h2-D3NS5-BCFT-p}).
The linking number $l^{(p)}=-N_{\rm D3}^{(p)}/N_{\rm D5}^{(p)}$ fixes the location of the flavors represented by the $p^{\rm th}$ D5-brane group in the quiver as $z_t=1-l^{(p)}/L$. This leads to
\be
i\pi (z_t-1)=i e^{\delta_p}k+\ln \frac{i e^{\delta_p}+1}{i e^{\delta_p}-1}~.
\ee
Upon exponentiating and acting with $W_k$ on both sides this leads to
\begin{align}
	W_k(-e^{-i\pi z_t})&=i e^{\delta_p} k~.
\end{align}
Using this in (\ref{eq:rho-t-sugra-balanced}) leads to the form in (\ref{eq:cR-v-gen-summary}).

In contrast to duals for 3d SCFTs, the exponentials in $h_{1/2}$ which give rise to the (locally) $\rm AdS_5\times S^5$ region, lead to $v$ being defined through a transcendental equation and to the generalized Lambert W-functions.
Moreover, only part of $\Sigma$ maps to eigenvalue distributions for BCFTs. Namely, the part explored by the D5$^\prime$ branes describing antisymmetric Wilson loops, which we used to derive  (\ref{eq:b-rho-h12}).
In fig.~\ref{fig:Sigma-h2D-contours} this is the part carved out by curves ending on the lower boundary of $\Sigma$ (see also \cite[sec.~5]{Coccia:2021lpp}).
In the $v$ coordinate in fig.~\ref{fig:BCFT-coord-transf} this part is $0\leq \Im(v)\leq\pi$. This means $t$ defined from $v$ via (\ref{eq:b-rho-h12}) has range $N_5$, as appropriate for a quiver with $N_5$ nodes.

\subsection{Examples}

In this section we work out the mapping to field theory parameters for a set of examples and give explicit expressions for the eigenvalue distributions.

\begin{figure}
	\includegraphics[width=0.4\linewidth]{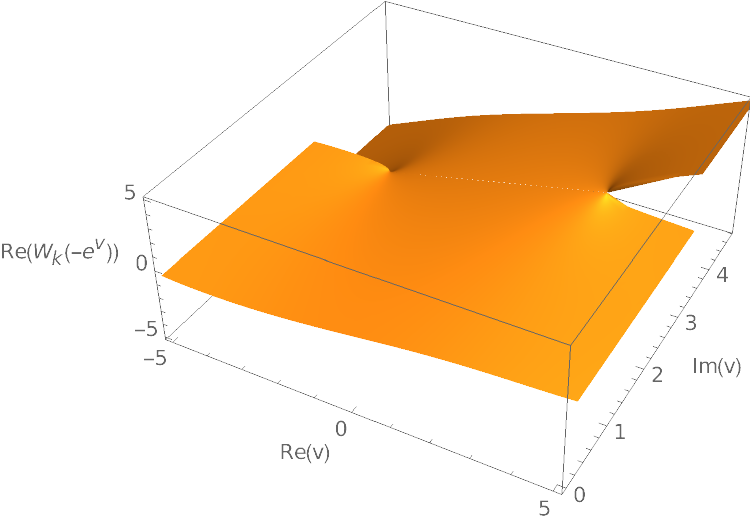}
	\hskip 5mm
	\includegraphics[width=0.4\linewidth]{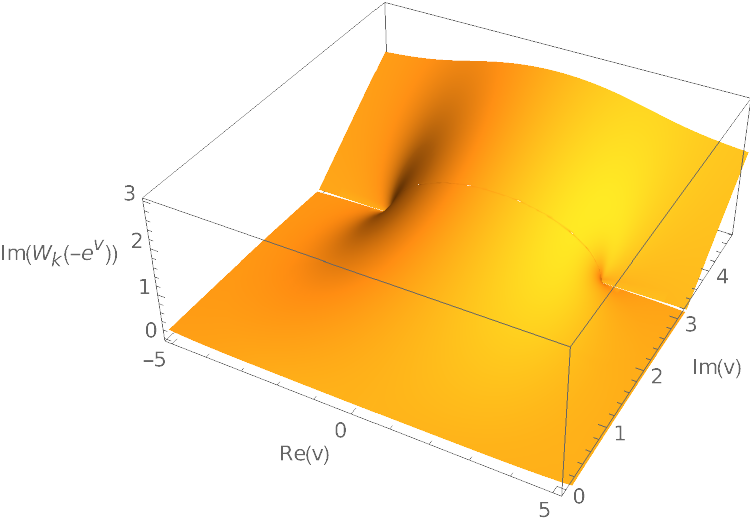}
\caption{Plots of the real (left) and imaginary (right) parts of $W_k(-e^v)$. The part relevant for the eigenvalue distributions is $0\leq\Im(v)\leq \pi$. Along the branch cuts $\mathcal B$ and $-\overline{\mathcal B}$ the imaginary part vanishes; the eigenvalue density at $z=1$ has compact support. For $0<z<1$ the support is non-compact. The real part is discontinuous across the cuts. For these plots we let $0\leq \Im(z)\leq \pi$, to cover the full eigenvalue densities.
\label{fig:Wk-plot}}
\end{figure}

The first BCFT is engineered by $N_5K$ D3-branes ending on $N_5$ NS5-branes, with $K$ D3-branes ending on each (fig.~\ref{fig:branes}). 
The quiver is in (\ref{eq:D3NS5-quiver})
and the harmonic functions specifying the holographic dual are given by (\ref{eq:h1h2-D3NS5-BCFT-p}) with $P=0$ (see also \cite[sec.~3.2]{Chaney:2024bgx}).
From the preceding discussion we find
\begin{align}
	\tilde\rho_t(\lambda)&=
	\frac{4N_5}{g_{4d}^2}\Im W_k(-e^{v})
	~,
	&
	v&=\frac{2\pi|\lambda|+i\pi t}{N_5}~.
\end{align}
Plots are shown in fig.~\ref{fig:Wk-plot}, where only $0\leq \Im(v)\leq \pi$ describes eigenvalue distributions.

The next example is the general solution with one group of D5-branes and one group of NS5-branes, i.e.\ $P=1$ in (\ref{eq:h1h2-D3NS5-BCFT-p}). To simplify notation, we set
\begin{align}
	N_{\rm D5}&\equiv N_{\rm D5}^{(1)}~, & \delta\equiv \delta_1~.
\end{align}
The numbers of D3-branes ending on the 5-brane groups can be computed following \cite[(4.27)]{Assel:2011xz}. For the D5 group at $z=\delta+\frac{i\pi}{2}$ and the NS5 group at $z=0$, the total D3-brane charges are
\begin{align}\label{eq:R-S}
	R\equiv \frac{\hat Q_{\rm D3}}{16\pi^2{\alpha'}^2 Q_{\rm NS5}}&=K+\frac{2\cot^{-1}\!e^\delta}{\pi}N_{\rm D5}~, 
	& 
	S\equiv\frac{Q_{\rm D3}}{16\pi^2{\alpha'}^2Q_{\rm D5}}&=Ke^{\delta+2\phi_0}-\frac{2\cot^{-1}\!e^\delta}{\pi}N_{5}~,
\end{align}
respectively. $R$ and $S$ give the net number of D3-branes ending on each 5-brane within the corresponding group (with sign for orientation). Special cases were used and studied in \cite{Uhlemann:2021nhu,Karch:2022rvr,Chaney:2024bgx,Coccia:2021lpp}.
The gauge theory description for the dual BCFT depends on the sign of $S$.
For $S>0$ the D5-branes impose partial Nahm pole boundary conditions on the 4d $\mathcal N=4$ SYM fields. For $S<0$ there is no Nahm pole and the quiver is as given in (\ref{eq:D5NS5K-quiver}).
The eigenvalue densities obtained from (\ref{eq:rho-t-sugra-balanced}) are
\begin{align}
	\tilde\rho_t(\lambda)&=
	\Im\left[
	\frac{4N_5}{g_{4d}^2}W_k(-e^{v})
	-\frac{i}{\pi} N_{\rm D5}\ln\left(\frac{W_k(-e^{v})-ik e^{\delta}}{W_k(-e^{v})+ike^{\delta}}\right)
	\right]
	~,
	&
	v&=\frac{2\pi|\lambda|+i\pi t}{N_5}~.
\end{align}
From (\ref{eq:R-S}) we find $ike^\delta=W_k(- e^{i\pi(1+S/N_5)})$. 
Noting that $1+S/N_5=T/N_5$, this leads to (\ref{eq:rhot-D3D5NS5-sum}). Comparing to (\ref{eq:D5NS5K-quiver}), this matches the general form in (\ref{eq:cR-v-gen-summary}), noting that the flavor node is at $z=T/N_5$.

As noted above, depending on the parameters in the supergravity solutions, the dual field theories can have partial Nahm pole boundary conditions. The transition here happens when $S$ turns positive and was discussed in \cite{Karch:2022rvr} (and for a more complicated example in \cite[sec.~2.2]{Deddo:2023oxn}).
While the field theory derivations in sec.~\ref{sec:matrix} only cover the case without Nahm pole, the supergravity derivation for the saddle points extends to cases with partial Nahm poles.

The discussion extends to general balanced quivers: Each flavor node corresponds to a D5-brane group in the supergravity solution. The locations of these groups, $\delta_p$, collectively determine the locations of the flavor nodes in the quiver and the saddle points can be read off from $h_1$.

\subsection{Unbalanced quivers}\label{sec:unbalanced}

We outline the generalization to BCFTs with 3d SCFTs on the boundary which have unbalanced nodes. This means there are multiple groups of NS5-branes where the number of D3-branes ending on each NS5 differs between the groups. In that case $h_2$ takes a more general form,
\begin{align}
	h_2&=\frac{\pi\alpha^\prime}{4}K e^{z+2\phi_0}-\frac{\alpha^\prime}{4}\sum_{q=1}^Q N_5^{(q)}\ln\tanh\left(\frac{z-\beta_q}{2}\right) +{\rm c.c.}
\end{align}
The total number of gauge nodes is $\sum_{q=1}^Q N_5^{(q)}$.
To find an explicit expression for the saddle point eigenvalue densities from (\ref{eq:b-rho-h12}) we again introduce a coordinate $v$ such that 
\begin{align}
	h_2&=\frac{\alpha'}{4}(v+\bar v)~.
\end{align}
This ansatz leads to 
\begin{align}
	\pi K e^{z+2\phi_0}-\sum_{q=1}^Q N_5^{(q)}\ln\tanh\left(\frac{z-\beta_q}{2}\right)&=v~.
\end{align}
We introduce $z=\ln(u/(\pi Ke^{2\phi_0}))$ and $\beta_q=\ln(\tilde\beta_q/(\pi Ke^{2\phi_0}))$, which leads to
\begin{align}
	u-\sum_{q=1}^Q N_5^{(q)}\ln\frac{u-\tilde\beta_q}{u+\tilde\beta_q}&=v~.
\end{align}
We note that the 5-brane numbers $N_5^{(q)}$ are quantized.
Inverting this relation leads to a further generalized version of the W-function. Namely  (see app.~\ref{sec:W} for notation)
\begin{align}
	u&=W\left({-\tilde \beta_1,\ldots, -\tilde\beta_1, -\tilde \beta_2,\ldots, -\tilde\beta_2,
		\cdots\cdots\cdots,-\tilde \beta_Q,\ldots, -\tilde\beta_Q
		\atop {+\tilde \beta_1,\ldots, +\tilde\beta_1, +\tilde \beta_2,\ldots, +\tilde\beta_2,
			\cdots\cdots\cdots,+\tilde \beta_Q,\ldots, +\tilde\beta_Q}};e^v\right)~,
\end{align}
where in the upper and lower rows of the first argument of $W$,  $\pm\tilde\beta_q$ is repeated $N_5^{(q)}$ times. The first two relations in (\ref{eq:b-rho-h12}) can now be solved to express $v$ in field theory variables. 
Substituting $z$ in $h_1$ then gives the saddle point eigenvalue densities through the last relation in (\ref{eq:b-rho-h12}), 
\begin{align}
	\tilde\rho_t(\lambda)&=\frac{2h_1(z,\bar z)}{\pi \alpha'}~,
	&
	z&=\ln\left[\frac{W\left({-\tilde \beta_1,\ldots, -\tilde\beta_1, -\tilde \beta_2,\ldots, -\tilde\beta_2,
			\cdots\cdots\cdots,-\tilde \beta_Q,\ldots, -\tilde\beta_Q
			\atop {+\tilde \beta_1,\ldots, +\tilde\beta_1, +\tilde \beta_2,\ldots, +\tilde\beta_2,
				\cdots\cdots\cdots,+\tilde \beta_Q,\ldots, +\tilde\beta_Q}};e^{2\pi\lambda+i\pi t}\right)}{\pi K e^{2\phi_0}}\right]~.
\end{align}
This can be connected to the previous formulation for $Q=1$ by noting  that $NW\left({\alpha \atop \beta};z\right)=W\left({\alpha/N, \ldots, \alpha/N \atop \beta/N, \ldots, \beta/N};z^N\right)$, where the ellipsis denote $N$-fold repetition.

\section{A precision test of top-down \texorpdfstring{A\lowercase{d}S/BCFT}{AdS/BCFT}}\label{sec:prec-text}

We showed that the harmonic functions describing the supergravity duals for BCFTs engineered by D3-branes ending on D5 and NS5 branes explicitly encode the saddle points dominating the field theory matrix models.
This relation was derived by equating field theory and gravity calculations of expectation values for Wilson loops in antisymmetric representations of individual gauge nodes in (\ref{eq:quiver-gen}). The Wilson loops extend along great circles on the boundary of the BCFT on $\rm HS^4$, and the combined expectation values for all gauge nodes and ranks of the representations contain sufficient information to reconstruct the entire family of saddle point eigenvalue densities for a given quiver.

Our results imply, by construction, that the string theory and field theory computations of antisymmetric Wilson loop expectation values agree. This extends the match between the two calculations, demonstrated in \cite{Coccia:2021lpp} for 3d SCFTs,\footnote{While \cite{Coccia:2021lpp} contained gravity calculations for BCFTs, the field theory computations were limited to 3d SCFTs.} to the entire class of BCFTs considered here. 
The string theory realization of the Wilson loops is through probe D5-branes for 3d SCFTs and BCFTs (and through fundamental strings for fundamental Wilson loops). The underlying brane setups differ in that there are semi-infinite D3-branes for BCFTs but not for 3d SCFTs, which translates to qualitatively different $h_{1/2}$, but the general results of \cite{Coccia:2021lpp} extend to BCFTs. With the saddle points in hand the field theory calculations can now also be extended to BCFTs. 
We thus obtain a host of new precision tests of the top-down AdS/BCFT dualities. In the following we illustrate this with explicit expressions for the BCFT without flavors in (\ref{eq:D3NS5-quiver}) as simplest example.\footnote{For this particular BCFT removing the semi-infinite D3-branes by setting $K=0$ trivializes the theory. For BCFTs with flavors analogous limits yield non-trivial 3d SCFTs, and the results for BCFTs reduce to those for 3d SCFTs.}

The simplest Wilson loop is the circular Wilson loop on the boundary of $\rm HS^4$ in the fundamental representation of the 4d gauge node in (\ref{eq:D3NS5-quiver}).
The expectation value, $\ln \langle W_f\rangle$, is determined in field theory at large $N$ by the largest eigenvalue $\lambda_{\rm max}$ associated with the 4d node, as $\ln \langle W_f\rangle=2\pi \lambda_{\rm max}$ (see e.g.\ \cite[sec.~3.2]{Coccia:2021lpp}). 
With $\lambda=Lx$ this becomes $\ln \langle W_f\rangle=2\pi L x_{\rm max}$ with $x_{\rm max}$ the upper bound of the support of the eigenvalue density at $z=1$, given below (\ref{eq:no-flavor-rho-norm}). This leads to
\begin{align}\label{eq:Wf}
	\ln \langle W_f\rangle&=L\left(\sqrt{k(k+2)}+\ln(k+1+\sqrt{k(k+2)})\right),
	&
	k&=\frac{\gym^2 K}{4N_5}~.
\end{align}
The parameter $k$ is order one while $L$ is large, so the entire expression is the leading-order result.
The same result can be obtained in string theory, where the Wilson loop is represented by a fundamental string embedded in the gravity dual. 
From the results in \cite[(4.12),(4.13)]{Coccia:2021lpp}, the BPS condition fixing the location of the string on $\Sigma$ is $\partial h_2=0$ on the boundary of $\Sigma$ where $h_1=0$, and the on-shell action is then given by $S_{\rm F1}=\frac{2}{\alpha'}|h_2|$. The expectation value of the fundamental Wilson loop is $\ln \langle W_f\rangle=S_{\rm F1}$. For $h_{1/2}$ in (\ref{eq:h1h2-D3NS5-BCFT-p}) this precisely reproduces the expression in (\ref{eq:Wf}).\footnote{%
For general balanced quivers as boundary degrees of freedom, $\ln\langle W_f\rangle$ takes the same functional form as in (\ref{eq:Wf}) but with modified $k$. For the theory in (\ref{eq:D5NS5K-quiver}) as example, $K$ is determined in terms of field theory data by (\ref{eq:RS-sum}).}

A similar analysis can be performed for Wilson loops in antisymmetric representations for all gauge nodes. The input is a choice of gauge node, parametrized as before by $z=t/L$, and the rank of the antisymmetric representation. We parametrize the rank as $\mathds{k} N(z)$ with $0\leq \mathds{k}\leq 1$ and denote the Wilson loop as $W_\wedge(z,\mathds{k})$.
The supergravity computations for the theory in (\ref{eq:D3NS5-quiver}) follow the logic detailed in \cite{Coccia:2021lpp}. Here we sketch the field theory calculation, starting with (see \cite[sec.~2.A]{Uhlemann:2020bek})
\begin{align}\label{eq:Wa-1}
	\exwil{\mk}&=-2\pi L \int_{b(z,0)}^{b(z,\mk)} dx~\vrho(z,x) x=-\frac{2L^3}{\pi \gym^2} \Im \int_{v_0}^{v_1}~dv~W_k(-e^{v})(v-i\pi z)~,
\end{align}
and, to define the function $b(z,\mk)$ in terms of $\mk$,
\begin{align}\label{eq:Wa-2}	
	\mk&=\frac{1}{N(z)}\int_{b(z,\mk)}^{b(z,0)} dx \vrho(z,x) =\frac{1}{2\pi k}\Im \int_{v_1}^{v_0} dv~W_k(-e^{v})~.
\end{align}
In both equations we used $\vrho$ from (\ref{eq:varrho-cR}), (\ref{eq: green source})
and introduced $v=2\pi x+i\pi z$ as before. The integrals over $v$ are along $\Im(v)=i\pi z$.
The lower bound is $v_0=\infty+i\pi z$ for $z<1$ and $v_0 =\sqrt{k(k+2)}+\ln(k+1+\sqrt{k(k+2)}) +i\pi$ for $z=1$, the upper bound is $v_1=2\pi b(z,\mk)+i\pi z$.

The integrals in (\ref{eq:Wa-1}) and (\ref{eq:Wa-2}) can be solved by change of variable.
We introduce a coordinate $u$ which matches a coordinate used in the holographic computations of \cite{Coccia:2021lpp}. It is defined as
\begin{align}\label{eq: u}
    u&=\frac{k-W_k(-e^{v})}{k+W_k(-e^{v})}~,&
    &\leftrightarrow & 
  v&=k\frac{1-u}{1+u}-\ln u~.
\end{align}
The lower integration bound becomes $u_0=0$ for $z<1$ and $u_0=-k-1+\sqrt{k (k+2)}$ for $z=1$. The integrals in (\ref{eq:Wa-1}), (\ref{eq:Wa-2}) become
\begin{align}\label{eq:Wa-1a}
\exwil{\mk} &= -\frac{2kL^3}{\pi \gym^2} \Im \int_{u_0}^{u_1} du \, \frac{ (u-1) \left( 2ku+(u+1)^2\right)}{u (u+1)^3} \left( k \frac{u-1}{u+1} + \ln u+ i \pi z\right)\,,
\\
\mk&=\frac{1}{2\pi }\Im \int_{u_1}^{u_0} du \frac{ (u-1) \left(2 k u+(u+1)^2\right)}{u (u+1)^3}\,,
\end{align}
where
\begin{align}\label{eq:u1-def}
	u_1&=\frac{k-W_k(-e^{2\pi b(z,\mk)+i\pi z})}{k+W_k(-e^{2\pi b(z,\mk)+i\pi z})}~.
\end{align}
Performing the integrals leads to
\begin{align}\label{eq:Wa-2-a}
		\exwil{\mk} 
	&= \frac{kL^3}{\pi \gym^2} \Im \Bigg(
	4 \operatorname{Li}_2(-u_1) +4\ln u_1 \ln(u_1+1)
	+ 2 i \pi z\left[ \ln\left(\frac{(u_1+1)^2}{u_1}\right)+k\frac{u_1^2 + 1}{(u_1+1)^2} \right]
	\nonumber\\
	&\hskip 22mm - \ln^2(u_1)+  2 k\frac{ u_1 ^2 + 1 }{(u_1+1)^2}\ln u_1    
	-k^2-k^2\frac{4+12u_1^2}{3 (u_1+1)^3}
	+\frac{4k}{u_1+1}
	\Bigg)~,
	\\ \label{eq:Wa-2-b}
	\mk&=
	\frac{z}{2}+\frac{1}{2\pi }\Im\left[\frac{2k u_1}{\left(u_1+1\right){}^2}-\ln\left(\frac{(u_1+1)^2}{u_1}\right)\right]~.
\end{align}
These expressions can be read in two ways: The first is to fix $z$ and $\mk$, i.e.\ gauge node and rank of the representation, then (\ref{eq:Wa-2-b}) with $u_1$ in (\ref{eq:u1-def}) determines $b$, which in turn determines the expectation value via (\ref{eq:Wa-2-a}).
The second option is to fix $z$ and $b$, which then directly determines $\mk$ via (\ref{eq:Wa-2-b}) and the expectation value via (\ref{eq:Wa-2-a}). Either way, the same expressions emerge from the supergravity computation following \cite{Coccia:2021lpp}, leading to a perfect match between holography and field theory.

The expressions simplify for the `maximal' $\mk=1/2$ Wilson loops. Due to the symmetry of the eigenvalue densities under $\lambda\rightarrow -\lambda$, $b(z,\frac{1}{2})=0$, leading to $u_1=(k-W_k(-e^{i\pi z}))/(k+W_k(-e^{i\pi z}))$. This can be used directly in (\ref{eq:Wa-1a}) to get the expectation values.

\let\oldaddcontentsline\addcontentsline
\renewcommand{\addcontentsline}[3]{}
\begin{acknowledgments}
	DH thanks Ming Yang for useful discussions.
	Part of this work was completed at the Aspen Center for Physics, which is supported by National Science Foundation grant PHY-2210452. DH is supported by FWO-Vlaanderen project G012222N, by the VUB Strategic Research Program High-Energy Physics and by a PhD fellowship from the VUB Research Council.
\end{acknowledgments}
\let\addcontentsline\oldaddcontentsline

\appendix

\section{Generalized Lambert W-function}\label{sec:W}

The standard Lambert W-function \cite{Corless1996OnTL} is defined as the inverse function of $f(w)=we^w$, i.e.\ by
\begin{align}
	we^w\big\vert_{w=W(z)}&=z~,
\end{align}
for complex $z$. For properties and references see \cite[§4.13]{NIST:DLMF}.
For the generalized W-function we follow the notation in \cite{2014arXiv1408.3999M}. That is, the generalized W-function is defined by
\begin{align}
	W\left({t_1,\ldots, t_n\atop {s_1,\ldots, s_m}};z\right):
	&&
	e^{w}\frac{(w-t_1)\dots (w-t_n)}{(w-s_1)\dots (w-s_m)}\Bigg\vert_{w=W\left({t_1,\ldots, t_n\atop {s_1,\ldots, s_m}};z\right)}&=z~.
\end{align}
With this definition $W\left(;z\right)=\ln z$ and $W\left({0\atop {\ }};z\right)=W(z)$ is the standard Lambert W-function.

\begin{figure}
\includegraphics[width=0.35\linewidth]{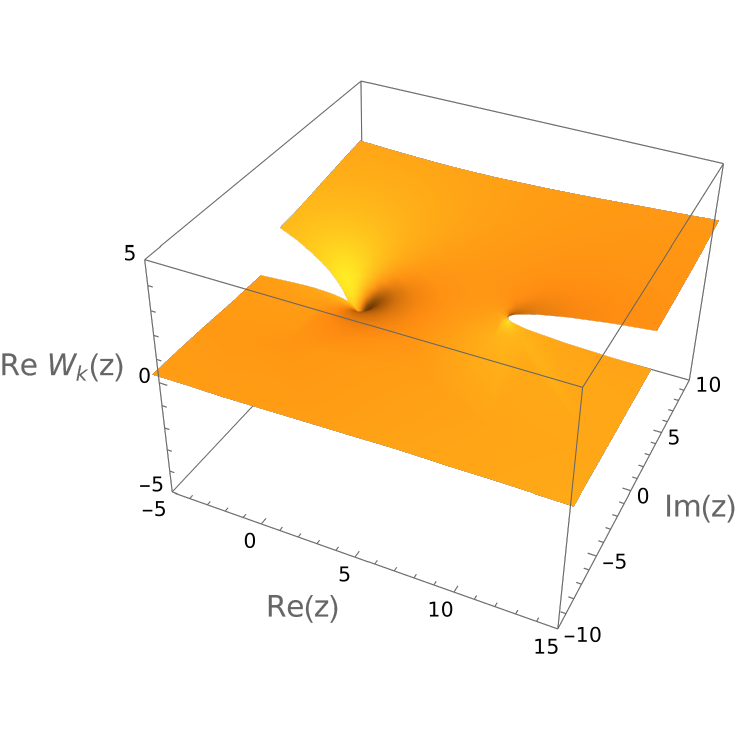}
\hskip 5mm
\includegraphics[width=0.35\linewidth]{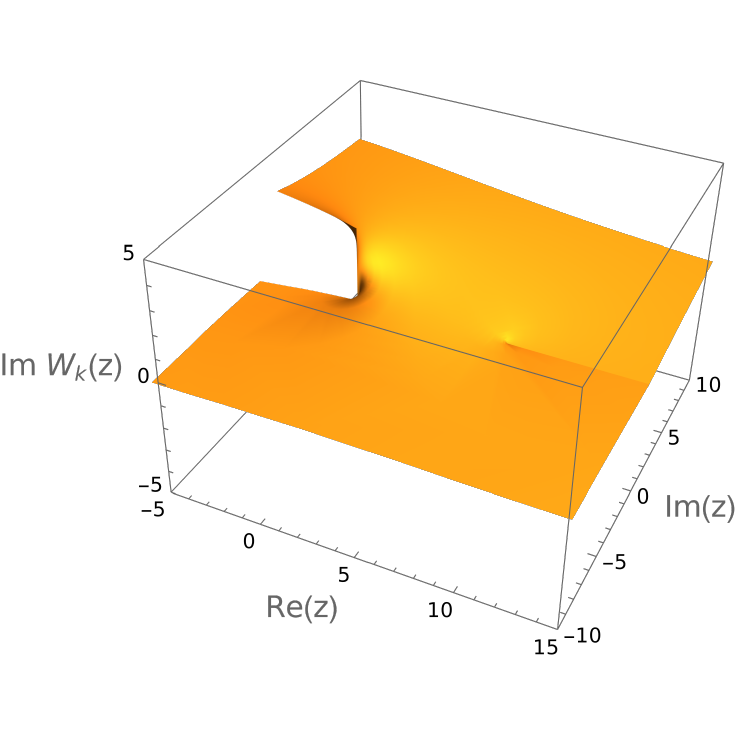}
\caption{Plots of $W_k(z)$, with the real part on the left and the imaginary part on the right, for $k=\frac{1}{2}$.\label{fig:Wk-plot-app}}
\end{figure}

A special case which we encounter frequently is $W_k(z)$, with $k>0$, defined as follows,
\begin{align}
W_k(z)&\equiv W\left({-k\atop {+k}};z\right): &
	e^{W_k(z)}\frac{W_k(z)+k}{W_k(z)-k}&=z~.&&
\end{align}
From the definition we have 
\begin{align}
	W_k(1/z)&=-W_k(z)~,
\end{align}
and
\begin{align}
    \partial_z W_k(z)&=e^{-W_k(z)}\frac{ \left(W_k(z)-k\right){}^2}{W_k(z){}^2-(k+2) k}
	=\frac{1}{z}\left(\frac{W_k(z){}^2-k^2}{W_k(z){}^2-(k+2) k}\right)~.
\end{align}
The points where the denominator vanishes are branch points, at $W_k(z)=\pm \sqrt{k(k+2)}$ or
\begin{align}
    z_{\pm}&=e^{\pm\sqrt{k(k+2)}}\frac{\sqrt{k(k+2)}\pm k}{\sqrt{k(k+2)}\mp k}~.
\end{align}
Both are positive, $z_+>z_-$ and $z_+z_-=1$.
We take the branch cuts outwards along the real axis.
An additional singularity is at $z=0$, where $\Re W_k\rightarrow-\infty$. It lies on the cut emerging from $z_-$ in the negative real direction.
We pick the branch of $W_k$ where for $z_-<\Re(z)<z_+$
\begin{align}\label{eq: w conjugate}
W_k(\overline z)&=\overline{W_k(z)}~.
\end{align}
A plot of this branch is shown in fig.~\ref{fig:Wk-plot-app}.
The imaginary part is continuous at $z_\pm$ and discontinuous at $z=0$.

\section{Supergravity derivation}\label{app:sugra-deriv}

We derive the relation between the harmonic functions $h_{1/2}$ specifying $\rm AdS_4\times S^2\times S^2\times\Sigma$ supergravity solutions and the saddle point eigenvalue distributions for the matrix models. The key idea is that the eigenvalue distributions can be extracted from the expectation values of Wilson loops in antisymmetric representations of individual gauge nodes, which can be computed in supergravity. 

The expectation value of a rank-$k$ antisymmetric Wilson loop,
denoted $\left\langle W_{\wedge}(z, \mathds{k})\right\rangle$ with normalized rank $\mathds{k}=k/N_t$ and $z=t/L$ labeling the gauge nodes, can be expressed as \cite{Coccia:2021lpp}
\begin{align}\label{eq:Wilson-app-0}
	\ln \left\langle W_{\wedge}(z, \mathds{k})\right\rangle&=2 \pi L N(z) \int_0^{\mathds{k}} dy\, b(z, y)~,
	&
	\int_{b(z, \mathds{k})}^{\infty} d x \hat{\rho}(z, x)=\mathds{k}~,	
\end{align}
where $\hat \rho$ is the unit-normalized eigenvalue distribution.
Since $\hat\rho$ is a non-negative function, $b(z,\mathds{k})$ is a monotonically decreasing function of $\mathds{k}$ at fixed $z$. As $\mathds{k}$ runs from 0 to 1, $b$ decreases from $+\infty$ to $-\infty$ (with obvious modification if $\hat\rho$ has compact support). The function $b$ at fixed $z$ is thus invertible. 
In supergravity, the Wilson loops are represented by probe D5 branes and the expectation values can be computed from their action, leading to
\begin{align} \label{eq: I int}
	\exwil{y}&=\frac{8}{\pi^2 \alpha'^3} \abs{I}~,
	&
	I&=\frac{1}{2}\int_{\xi_0}^{\xi_1(z,\mk)}d\xi\, h_1 h_2 \p_{\xi} h_2~,
\end{align}
where the integral is along a countour of constant $h_2^D$.
The starting point is fixed, while the end point $\xi_1(z,\mathds{k})$ depends on the rank of the representation and the gauge node.
The gauge node $z$ is extracted from D3-brane charge $N_{\rm D3}$ carried by the probe D5, and the rank of the representation $\mathds{k}$ from its F1 charge $N_{\rm F1}$. They are
\begin{align}\label{eq: n' int}
	N_{\rm D3}&=\frac{2}{\pi \alpha'}h_2^D~, &
	N_{\rm F1}&=\frac{4}{\pi^2 \alpha'^2} N'~,
	\qquad
	N'=\frac{1}{2}\int_{\xi_0}^{\xi_1(\mk)} h_1 \p_{\xi} h_2~.
\end{align}
with $z L=\abs{N_{\rm D3}}$ and $\mk N(z)=\abs{N_{\rm F1}}\mod{N(z)}$.
The Wilson loop expectation values are invariant under $\mathds{k}\rightarrow 1-\mathds{k}$. 
The corresponding probe D5-branes wrap an equatorial $S^1$ in one of the $S^2$'s at $\xi_0$; this $S^1$ slips along the embedding and collapses at a pole at $\xi_1$. Exchanging the poles, with the embedding otherwise unchanged, corresponds to $\mathds{k}\rightarrow 1-\mathds{k}$. The full collection of Wilson loops can thus be identified with a double cover of $\Sigma$ through the end points of the D5 embeddings.

To explicitly extract the eigenvalue distributions, we start by taking derivatives of the two equations in (\ref{eq:Wilson-app-0}) at fixed $z$. We can extract the normalized eigenvalues and distributions as
\bea \label{eq: b rho int}
b(z,\mathds{k})&=\frac{1}{2 \pi LN(z)}\frac{d\ln \left\langle W_{\wedge}(z, \mathds{k})\right\rangle}{d\mathds{k}}~,\\
\hrho(z,x)&=-\deri{\mathds{k}}{x}=-2 \pi LN(z)(\derit{\exwil{\mathds{k}}}{\mathds{k}})^{-1}\Big\vert_{\mathds{k}=b^{-1}(z,x)}~.
\eea
We then proceed with similar derivatives of the supergravity expressions,
\begin{align} \label{eq: I N'int}
\deri{I}{N'}&=\frac{dI/d\mathds{k}}{dN'/d\mathds{k}}=h_2(\xi_1(\mk))~,
&
\deri{N'}{h_2}&=\frac{dN'/d\mathds{k}}{dh_2/d\mathds{k}}=\frac{1}{2}h_1(\xi_1(\mk))~.
\end{align}
Comparing \eqref{eq: I N'int} with \eqref{eq: b rho int}, we have
\bea
\deri{\exwil{\mk}}{N_{F1}}&=\frac{2}{\alpha'}\deri{I}{N'}=\frac{2}{\alpha'} h_2(z,\mk)~,\\
\derit{\exwil{\mk}}{N_{F1}}&=-\frac{\pi^2 \alpha'}{h_1(z,\mk)}~.
\eea
Collecting the coefficients in the definition of Wilson loops and F1 charge, we have our desired relations
\bea\label{eq: rho h1 b h2}
\hrho(z,x) &=\frac{2 L}{N(z) \pi \alpha'} h_1(z,\mk)\Big\vert_{\mathds{k}=b^{-1}(z,x)},\\
b(z,\mk)&=\frac{h_2(z,\mk)}{\pi L \alpha'}.
\eea
These expressions lead to (\ref{eq:b-rho-h12}).
For regular solutions $h_{1/2}$ are non-negative, but they cover the full range of eigenvalue densities, see footnote \ref{foot:ext}.

\bibliography{BCFT}
\end{document}